%Paper: hep-th/9402092
%From: Masayuki Matsuzaki <matsuza@fukuoka-edu.ac.jp>
%Date: Wed, 16 Feb 94 19:42:31 +0900

%%%%%%%%%%%%%%%%%%%%%%%%%%%%%%%%%%%%%%%%%%%%%%%%%%%%%%%%%%%%%%%%%
%                                                               %
%     A set of macros for preparing a manuscript of a paper     %
%                                                               %
%       Type <\input tmacro> at the top of your text file       %
%                                                               %
%                  December 1992,   K. Takada                   %
%                                                               %
%%%%%%%%%%%%%%%%%%%%%%%%%%%%%%%%%%%%%%%%%%%%%%%%%%%%%%%%%%%%%%%%%
%
%  ********** Setting dimensions and style of document **********
%
\magnification=\magstep1
\hsize=15.5 true cm
\vsize=22.0 true cm
%\hoffset=0.5 true cm
%\voffset=1.0 true cm
\baselineskip=0.86 true cm
\parindent=30pt
\footline={\hfill --\folio -- \hfill}
%
%  ********** Definition of fourteen-point fonts ****************
%

\font\fourteeni=cmmi10 scaled\magstep2	    \skewchar\fourteeni='177
\font\fourteensy=cmsy10 scaled\magstep2	    \skewchar\fourteensy='60

%
%  ********** Definition of twelve-point capital fonts **********
%

%
%  ********** Definition of ten-point capital fonts *************
%

%
%  ********** Definition of math-boldface fonts family **********
%
\font\tenmib=cmmib10			    \skewchar\tenmib='177
\newfam\mibfam
\def\tenpoint{\relax
    \def\mib{\fam\mibfam \tenmib}
    \textfont\mibfam=\tenmib	    \scriptfont\mibfam=\tenmib
    \scriptscriptfont\mibfam=\tenmib}
\tenpoint

%
%  ********** Definition of fonts of Japanese characters ********
%

%
%  ********** Change of command for gothic letters **************
%

%
%  ********** Roman numerals ************************************
%
%       ***** upper case Roman numerals *****
\def\Roman#1{\uppercase\expandafter{\romannumeral#1}}
\def\ucrn#1{\uppercase\expandafter{\romannumeral#1}}
\def\rnuc#1{\uppercase\expandafter{\romannumeral#1}}
%       ***** lower case Roman numerals *****
\def\roman#1{\lowercase\expandafter{\romannumeral#1}}
\def\lcrn#1{\lowercase\expandafter{\romannumeral#1}}
\def\rnlc#1{\lowercase\expandafter{\romannumeral#1}}
%
%  ********** Easy macros for Greek letters *********************
%

%
%  ********** Boldface Greek Letters ****************************
%

%
%  **************************************************************
%

%
%  ********** Other useful macros *******************************
%
\def\frac#1#2{{{#1}\over{#2}}}
\def\fract#1/#2{\leavevmode\kern.1em
\raise.5ex\hbox{\the\scriptfont0 #1}\kern-.1em
/\kern-.15em\lower.25ex\hbox{\the\scriptfont0 #2}\thinspace}
\def\dg{\dagger}

%
%  ********** Yen mark ******************************************
%
\def\up#1#2{\smash{\raise#2\hbox{#1}}}

%
%  **************************************************************
%
\def\l"{\char'134}
\def\r"{\char'42}
%
%  ********** Names of journals *********************************
%

%
%  ********** Boson mappings ************************************
%

%
%  ********** CG coefficient ************************************
%
\def\clebsch(#1,#2,#3,#4,#5,#6){
    \langle #1 #2 #3 #4 \vert #5 #6 \rangle\thinspace }
%
%  ********** 6-j symbol ****************************************
%
\def\sixj(#1,#2,#3,#4,#5,#6){
    \left\{ \matrix { #1 & #2 & #5 \cr
                      #4 & #3 & #6 \cr } \right\} }
%
%  ********** 9-j symbol ****************************************
%
\def\ninej(#1,#2,#3,#4,#5,#6,#7,#8,#9){
    \left\{ \matrix { #1 & #2 & #3 \cr
                      #4 & #5 & #6 \cr
                      #7 & #8 & #9 \cr } \right\} }
%
%  ********** Limit symbol **************************************
%
\def\limit#1#2{\enskip
\hbox to 28pt{$\,$\rightarrowfill$\,$}\hskip -28pt
_{\lower4pt\hbox to 28pt{\hfill$\scriptstyle #1\rightarrow#2$\hfill}}
\enskip}
%
%             *********************
%

%
%  ********** Commutator Bracket ( [ , ] ) **********************
%
\def\com[{\hbox{$\lceil\!\!\!\hskip 0.18mm\lfloor$}}
\def\com]{\hbox{$\rceil\!\!\!\hskip 0.18mm\rfloor$}}
%
%  ********** Position of Figure Insertion **********************
%
%  This is the definition of macro to make
%
%            +--------------+
%            |              |
%            |    Fig. #1   |
%            |              |
%            +--------------+
%

%
%  ********** Position of Table Insertion ***********************
%
%  This is the definition of macro to make
%
%            +---------------+
%            |               |
%            |    Table #1   |
%            |               |
%            +---------------+
%

%
%  ********** Position-raising or lowering macros ***************
%
%       Raise #1 by #2; e.g. \up{ABC}{2.5ex} raises ABC by 2.5ex.
%       Lower #1 by #2; e.g. \down{XYZ}{0.5cm} lowers XYZ by 5 mm.
%
\def\up#1#2{\smash{\raise#2\hbox{#1}}}
\def\down#1#2{\smash{\lower#2\hbox{#1}}}
%
%  ********** New type of \geq and \leq ( \geqq, \leqq ) ********
%

%
%  ********** Line spacing **************************************
%

%
% ***************************************************************

\font\threebf=cmbx10 scaled\magstep3

	    \skewchar\fourteeni='177
	    \skewchar\fourteeni='177

%
%  **************************************************************
%

%
%  **************************************************************
%

%-------------------------------------------------

%
%  **************************************************************
%

%
%  **************************************************************
%

\baselineskip=0.9 true cm

\pageno=1

%%refnostyle = PR
%
%
%
%
%
%
\vskip 1.0 true cm
\centerline{\threebf Two-dimensional SU(N) Gauge Theory }
\vskip 0.3 true cm
\centerline{\threebf on the Light Cone}

\vskip 1.5 true cm
\centerline{Takanori Sugihara}
\centerline{\it Department of Physics, Kyushu University,
                                       Fukuoka 812, Japan}
\vskip 1.0 true cm
\centerline{Masayuki Matsuzaki}
\centerline{\it Department of Physics, Fukuoka University of Education,
                         Munakata, Fukuoka 811-41, Japan}
\vskip 1.0 true cm
\centerline{Masanobu Yahiro}
\centerline{\it University of Fisheries, Shimonoseki 759-65, Japan}
\vskip 1.0 true cm

     Two-dimensional SU($N$) gauge theory is accurately analyzed
with the light-front Tamm-Dancoff approximation, both numerically and
analytically. The light-front Einstein-Schr\"odinger equation
for mesonic mass reduces to the 't Hooft equation
in the large $N$ limit, $g^2N$ fixed, where $g$ is the coupling constant.
Hadronic masses are numerically obtained
in the region of $m^2 \ll g^2N$, where $m$ is the bare quark
(q) mass. The lightest mesonic and baryonic states are
almost in valence.  The second lightest mesonic state is
highly relativistic in the sense that
it has a large 4-body ($ {\rm qq} \bar{\rm q} \bar{\rm q} $) component
in addition to the valence (${\rm q} \bar{\rm q}$) one.
In the strong coupling limit our results are consistent with
the prediction of the bosonization for ratios of
the lightest and second lightest mesonic masses to the
lightest baryonic one. Analytic solutions to the lightest hadronic masses
are obtained, with a reasonable approximation, as
$\sqrt{2Cm}(1-1/N^2)^{1/4}$
in the mesonic case and $\sqrt{CmN(N-1)}(1-1/N^2)^{1/4}$
in the baryonic case,
where $C=(g^2N\pi/6)^{1/2}$. The solutions well reproduce
the numerical ones. The $N$- and $m$-dependences
of the hadronic masses are explicitly shown by the analytical solutions.

\vfill\eject

\leftline{\bf 1. INTRODUCTION AND SUMMARY}
\vskip 1.0 em

    Two-dimensional SU($N$) quantum chromodynamics (${\rm QCD}(N)_2$)
is a good model for studying ideas and tools which might be feasible
in analyses of QCD in 3+1 dimensions.
't Hooft introduced the model to test the power of
the $1/N$ expansion[1].   He summed planar diagrams
which dominate the leading order in the expansion
and derived an equation. The 't Hooft equation is valid
in the large $N$ limit, $g^2N$ fixed, where $g$ is
the coupling constant. The mass spectrum of the equation
reveals a nearly straight "Regge trajectory".

     The large $N$ limit corresponds to the weak coupling one, since
$g^2N$ is fixed. The $1/N$ expansion then works
in the weak coupling regime ($g \ll m$, $m$ being the bare quark mass),
but not in the strong coupling one, because
it is almost impossible to calculate higher-order terms
in the expansion.  For this reason, QCD$(N)_2$ in the strong coupling
regime has been studied with some other methods so far.
Nevertheless, the dynamics is not understood well in the region.
The bosonization predicts ratios of meson masses to a baryon mass[2],
but they are valid only in the strong coupling limit.
The lattice calculation has given a low-energy spectrum of SU(2),
but their accuracy is very poor [3].

    The discretized light-cone quantization (DLCQ) has been proposed as
a useful tool for computing the hadronic mass[4$-$8].
The mass obtained with
DLCQ is a function of $K$, the parameter which characterizes
the discretization of the total light-cone momentum, $\cal P$.
One then has to take the large $K$ limit
to get a physical mass containing
no unphysical parameter, but the convergence is very slow
for large $g$[5]. Increasing $K$ demands a lot
of numerical efforts, so that a reasonably large $K$ was not taken
in calculations done so far for strong coupling[6].

    Recently the light-front Tamm-Dancoff (LFTD) approximation [9]
has been
proposed as one of the alternative non-perturbative tools
to the lattice gauge theory. In the standard equal-time field theory
the vacuum state is a sea of an infinite number of constituents
( quarks (q) and gluons in QCD ) and it is then unlikely that the vacuum
and hadronic states are well described with a finite number
of constituents. In fact such a truncation of the Fock space,
i.e. the Tamm-Dancoff approximation[10], causes
some serious problems[9]. Such problems do not appear in light-front
field theory[11],
owing to the fact that the vacuum on the light cone is trivial[9].
The LFTD approximation is the Tamm-Dancoff approximation applied to
light-front field theory.
Both LFTD and DLCQ are based on light-front field theory, but
LFTD might be more reliable than DLCQ for strong coupling [12].

      On the other hand, LFTD includes its own problems; (a) Non-perturbative
renormalization, (b) the relation between spontaneous symmetry breaking
and the triviality of the vacuum ( what is called, the "zero-mode" problem )
and (c) recovery of rotational symmetry. These essential problems
has not been settled yet.
They, however, would not appear in two-dimensional models
such as QCD$(N)_2$.  Coleman's theorem[13]
says that symmetry breaking of global
continuous symmetry does not occur in two dimensions. We then do not face
the problem (b). The problem (a) and (c) do not exist, from the outset,
in two dimensions.

    In this paper, we study ${\rm QCD}(N)_2$ in the region of $g^2N \gg m^2$
with LFTD; the region corresponds to the strong coupling
region ( $g \gg m$ ) for small $N$, and for large $N$ it covers not only
the strong coupling region but also the medium and weak coupling ones.
We first derive the light-cone Hamiltonian, $P^-$, and the
Einstein-Schr\"odinger (ES) equation,
$2{\cal P}P^-|\Psi>=M^2|\Psi>$,
for hadronic mass and wave function, $M$ and $\Psi$, in the
framework of light-front field theory [11].
As the Tamm-Dancoff approximation,
the mesonic wave function with SU($N$) symmetry is truncated to 2-body
( ${\rm q}\bar {\rm q}$ ) and 4-body ( ${\rm qq}\bar {\rm q}\bar {\rm q}$ )
components, and the baryonic state to the $N$-body (${\rm q}^N$) one.
Inclusion of the 4-body state is essential to obtain the results
(ii) and (iii) mentioned in the next paragraph.
The ES equation is numerically solved by diagonalizing $P^-$
within a space spanned by a finite number of
basis functions. All tools needed for this calculation are prepared by
our previous work [14] for the massive Schwinger model.

Our main results are summarized as follows.

\noindent (i) $P^-$ involves a term proportional to
$Q^2/\eta$, where $Q^2$ is the Casimir operator of SU($N$)
and $\eta$ is an infinitesimal constant. The term enforces confinement,
restricting finite energy solutions to color singlets. This is a well-known
property of ${\rm QCD}(N)_2$.

\noindent (ii) The 't Hooft equation is compared with the ES equation which
 couples
the 2-body wave function with the 4-body ones. The couplings are of order
$1/\sqrt{N}$, so that the 2-body sector of the ES equation is decoupled from
the 4-body one in the large $N$ limit. The 2-body sector, as expected,
reduces to the 't Hooft equation in the limit. The 4-body sector does not
generates any bound state in the limit.

\noindent (iii) Numerical solutions of the ES equation yield two mesonic and
one baryonic bound states for small $N$. Masses of the three states
tend to zero as $m^{0.5}$ in the massless limit
( $m \rightarrow 0$ ), as expected from PCAC [14].
Ratios of the two mesonic
masses to the baryonic one at small $m$ consist
with the predictions [2] of the
bosonization within $\sim 7\%$ error.
The lowest mesonic and baryonic states are almost in valence.
The second mesonic state is highly
relativistic in the sense that it has the ${\rm q}\bar {\rm q}$ and
${\rm qq}\bar {\rm q}\bar {\rm q}$ components with almost the same
magnitude.
The existence of such a relativistic state is unpredictable from
the diagramatic consideration based on the $1/N$ expansion, since
the 't Hooft equation as a result of the consideration generates
only 2-body states.

\noindent (iv) Assuming that the lightest mesonic and baryonic states are
in valence, we can obtain approximate solutions to their masses as
$\sqrt{2Cm}(1-1/N^2)^{1/4}$
in the mesonic case and $\sqrt{CmN(N-1)}(1-1/N^2)^{1/4}$
in the baryonic case,
where $C=(g^2N\pi/6)^{1/2}$. The assumption assures that
the approximate solutions are accurate
in the region of $g^2N \gg m^2$. They show
the $N$-dependence of the masses explicitly
in the whole range of $N$. The leading order in $1/N$
is $O(N^0)$ for the mesonic mass and $O(N)$ for the baryonic one,
as predicted by some works [15$-$17] based on the expansion.
A theoretical surprise is that the next-to-leading order is
not $O(N^{-1})$ but $O(N^{-2})$ for the mesonic mass.
This makes the 't Hooft solution more reliable in the case
of the lightest mesonic mass.

     We derive the light-cone Hamiltonian and the result (i)
in sec. 2.1 and the ES equation for hadronic mass and the result (ii)
in sec. 2.3. In sec. 2.2 the color-singlet states of meson and baryon
are constructed within the truncated Fock space.
We first present, in sec. 2.4, the approximate solutions
to the lightest hadronic masses and the result (iv).
The approximate but analytic solutions are
convenient to see the $N$- and $m$-dependences of the masses explicitly.
Accuracy of the approximate solutions are tested in sec. 3.
Numerical methods and the result (iii) are also presented there.
Section 4 is devoted to discussions. Appendices are collections of
lengthy expressions.

\vskip 2.0 em
\leftline{\bf2. LIGHT-FRONT TAMM-DANCOFF APPROXIMATION}
\vskip 1.0 em
\leftline{2.1. Light-cone Hamiltonian}
\vskip 1.0 em

%%eqnostart = (2.1)

     The Lagrangian density of ${\rm QCD}(N)_2$
for interacting quark and gauge fields, $\psi$ and $A_{\mu}^a$
($a=1$ to $N^2-1$), is
	$$
	{\cal L}=-\frac{1}{4}{F_{\mu\nu}^{a}F^{a\mu\nu}}
	+\bar\psi(i\gamma^{\mu}D_{\mu}-m)\psi,                 \eqno{\rm (2.1)}
	$$
where $
	D_{\mu}=\partial_{\mu}-igA_{\mu}^{a}T^{a}
      $
and
      $
    F_{\mu\nu}^{a}=\partial_{\mu}A_{\nu}^{a}-\partial_{\nu}A_{\mu}^{a}
	+gf_{abc}A_{\mu}^{b}A_{\nu}^{c}
      $
for the generator $T^a$ and the structure constant $f_{abc}$
of SU($N$).
Light-front field theory [11] starts with the
introduction of light-cone
coordinates, $x^{\mu}=(x^+, x^-)\equiv ((x^0+x^1)/\sqrt{2},
(x^0-x^1)/\sqrt{2})$; for any other vector, $V^{\pm}=(V^0 \pm V^1)/\sqrt{2}$.
( We take the same notations and conventions as in Ref. [14].)
The equations of motion are
$$\eqalign{
	i\sqrt{2}\partial_{-}\psi_{\rm L}= & m\psi_{\rm R},\cr
	i\sqrt{2}\partial_{+}\psi_{\rm R}= & m\psi_{\rm L}
	-\sqrt{2}gA^{-}\psi_{\rm R},                  \cr
	\partial_{-}^{2}A^{a-}= & \sqrt{2}g{\psi_{\rm R}}^{\dg}
             T^{a}\psi_{\rm R},       \cr
        -\partial_{-}\partial_{+}A^{a-}
	= & \sqrt{2}g{\psi_{\rm L}}^{\dg}T^{a}\psi_{\rm L}
	+gf_{abc}A^{b-}\partial_{-}A^{c-}.       \cr}          \eqno{\rm (2.2)}
$$
for the light-cone gauge, $A^{a+}=0$, where
$\psi=(\psi_{\rm R},\psi_{\rm L})^T$.
The first and third equations do not involve the time derivative
($\partial_{+}$) and are therefore just constraints which determine
$\psi_{\rm L}$ and $A^-$ in terms of $\psi_{\rm R}$. Thus, $\psi_{\rm L}$
and $A^-$ are
not independent variables and not subject to a quantization condition.
The constraints are then solved with the inverse derivative operator
$\partial_{-}^{-1}$,
	$$\eqalign{\hskip 1.5 true cm
        \psi_{\rm L}(x^{-})
      = & -i\frac{m}{\sqrt{2}}\frac{1}{\partial_{-}}\psi_{\rm R} \cr
      = & -i\frac{m}{2\sqrt{2}}
	                \int dy^{-}\epsilon(x^{-}-y^{-})
			\psi_{\rm R}(y^{-}),             \cr}  \eqno{\rm (2.3)}
	$$
	$$
	A^{a-}(x^{-})
      =  \sqrt{2}g\frac{1}{\partial_{-}^2}
           {\psi_{\rm R}}^{\dg}(x^{-})T^{a}\psi_{\rm R}(x^{-}),\eqno{\rm (2.4)}
	$$
where $\epsilon(x)$ is $1$ for $x>0$ and $-1$ for $x<0$. The only independent
variable $\psi_{\rm R}$ is quantized by an anticommutation relation at the
equal light-cone time $x^+=y^+$,
	$$
	{ \{ \psi_{i{\rm R}}(x),{\psi_{j{\rm R}}}^{\dg}(y) \} }_{x^{+}=y^{+}}
	=\frac{1}{\sqrt{2}}\delta_{ij}\delta(x^{-}-y^{-}).    \eqno{\rm (2.5)}
	$$
The use of the light-cone coordinates and light-cone gauge thus reduces
the number of independent variables. This is an advantage of
light-front field theory. The energy-momentum vectors commute mutually
and are therefore the constants of motion. The time component
( light-cone Hamiltonian ) is
	$$\eqalign{
	 P^{-}
       = & -\frac{im^{2}}{2\sqrt{2}}\int dx^{-}dy^{-}
      {\psi_{\rm R}}^{\dg}(x^{-})\epsilon(x^{-}-y^{-})\psi_{\rm R}(y^{-})  \cr
         & -\frac{g^{2}}{2}\int dx^{-}
	    j^{a+}(x^{-})\frac{1}{\partial_{-}^{2}}
	    j^{a+}(x^{-}),  \cr}
                                                               \eqno{\rm (2.6)}
	$$
and the spatial one ( light-cone momentum ) is
	$$
	P^{+}=i\sqrt{2}\int dx^{-}{\psi_{\rm R}}^{\dg}(x^{-})
	      \partial_{-}\psi_{\rm R}(x^{-}).                 \eqno{\rm (2.7)}
	$$
The field $\psi_{\rm R}$ is expanded at $x^+=0$ in terms of free waves [18],
each with momentum $k^+$,
	$$
	\psi_{i{\rm R}}(x^{-})=\frac{1}{2^{1/4}}\int_{0}^\infty
	\frac{dk^{+}}{2\pi\sqrt{k^{+}}}
	[b_{i}(k^{+})e^{-ik^{+}x^{-}}
	+{d_{i}}^{\dg}(k^{+})e^{ik^{+}x^{-}}],                 \eqno{\rm (2.8)}
	$$
with
	$$
	 \{ b_{i}(k^{+}),{b_{j}}^{\dg}(l^{+}) \}
         =\{ d_{i}(k^{+}),{d_{j}}^{\dg}(l^{+}) \}
         =2\pi k^{+}\delta_{ij}\delta(k^{+}-l^{+}),            \eqno{\rm (2.9)}
	$$
The color current,
$j^{a+} \equiv \sqrt{2}:{\psi_{\rm R}}^{\dg}T^{a}\psi_{\rm R}:$,
is normal-ordered with respect to the creation and annihilation operators.
The charge is then
	$$
	Q^{a}=\int dx^{-}j^{a+}
         =\sum_{i,j}^{N}(T^{a})_{ij}
	      \int_{0}^{\infty}\frac{dk^{+}}{2\pi k^{+}}
	      [{b_{i}}^{\dg}(k^{+})b_{j}(k^{+})
	      -{d_{j}}^{\dg}(k^{+})d_{i}(k^{+})].             \eqno{\rm (2.10)}
	$$
The last term in $P^-$ can be rewritten
with the standard Fourier transform [19],
	$$\eqalign{
	& \int dx_{1}^{-}
	    j^{a+}(x_{1}^{-})\frac{1}{\partial_{-}^{2}}
	j^{a+}(x_{1}^{-}),                                \cr
	= & \frac{1}{4}\int dx_{1}^{-}dx_{2}^{-}dy^{-}
	 j^{a+}(x_{1}^{-})\epsilon(x_{1}^{-}-y^{-})
	 \epsilon(y^{-}-x_{2}^{-})j^{a+}(x_{2}^{-}),      \cr
	= & \frac{1}{2}\int dx_{1}^{-}dx_{2}^{-}
	 j^{a+}(x_{1}^{-})|x_{1}^{-}-x_{2}^{-}|j^{a+}(x_{2}^{-})
	-\frac{1}{4\eta}\sum_{a=1}^{N^{2}-1}Q^{a}Q^{a}
        +O(\eta),                                     \cr}    \eqno{\rm (2.11)}
	$$
where use has been made of
	$$
	\epsilon(x)=\frac{1}{2\pi i}\int_{-\infty}^{\infty}dk\bigg(
	            \frac{1}{k+i\eta}+\frac{1}{k-i\eta}
	            \bigg) e^{ikx},                           \eqno{\rm (2.12)}
	$$
	$$
	|x|=\frac{1}{2\pi i^{2}}\int_{-\infty}^{\infty}dk\bigg[
	    \frac{1}{(k+i\eta)^{2}}+\frac{1}{(k-i\eta)^{2}}
	    \bigg] e^{ikx}.                                   \eqno{\rm (2.13)}
	$$
The term $Q^{2}/4\eta$ ($Q^2 \equiv \sum Q^{a}Q^{a}$)
enforces confinement, restricting finite eigenvalues to the color-singlet
( $Q^2=0$ ) subspace.

     The Hamiltonian is expressed with the creation and annihilation
operators,
        $$
	P^{-}=P^{-}_{\rm free}+P^{-}_{\rm self}+P^{-}_{0}+P^{-}_{2},
\eqno{\rm (2.14)}
        $$
	$$\eqalign{
	P^{-}_{\rm free}= & \frac{m^{2}}{4\pi}\sum_{i=1}^{N}\int_{0}^{\infty}
	\frac{dk}{k^{2}}
	[{b_{i}}^{\dg}(k)b_{i}(k)+{d_{i}}^{\dg}(k)d_{i}(k)],             \cr
	P^{-}_{\rm self}= & \frac{g^{2}}{8\pi^{2}}\sum_{a=1}^{N^{2}-1}
        \sum_{i,j,k,l}^{N}(T^{a})_{ij}(T^{a})_{kl}
	\int_{0}^{\infty}\frac{dk_{1}}{k_{1}}\delta_{jk}
	[{b_{i}}^{\dg}(k_{1})b_{l}(k_{1})+{d_{l}}^{\dg}(k_{1})d_{i}(k_{1})]\cr
	&\times\int_{0}^{\infty}dk_{2}
	\bigg[\frac{1}{(k_{1}-k_{2})^{2}}
       -\frac{1}{(k_{1}+k_{2})^{2}}\bigg],                               \cr
	P^{-}_{0}= & \frac{g^{2}}{8\pi^{3}}\sum_{a=1}^{N^{2}-1}
        \sum_{i,j,k,l}^{N}(T^{a})_{ij}(T^{a})_{kl}
	\int_{0}^{\infty}\prod_{m=1}^{4}
	\frac{dk_{m}}{\sqrt{k_{m}}}
	\delta(k_{1}+k_{2}-k_{3}-k_{4})                                  \cr
	& \times \biggl\{ \big[{b_{i}}^{\dg}(k_{1}){b_{k}}^{\dg}(k_{2})
			        b_{l}       (k_{3}) b_{j}       (k_{4})
			      +{d_{j}}^{\dg}(k_{1}){d_{l}}^{\dg}(k_{2})
			        d_{k}       (k_{3}) d_{i}       (k_{4})
	\big] \frac{1}{2(k_{1}-k_{4})^{2}}                               \cr
	                    -{b_{i}}^{\dg}(k_{1})&{d_{l}}^{\dg}(k_{2})
			      d_{k}       (k_{3}) b_{j}       (k_{4})
			     \frac{1}{(k_{1}-k_{4})^{2}}
			    +{b_{i}}^{\dg}(k_{1}){d_{j}}^{\dg}(k_{2})
			      d_{k}       (k_{3}) b_{l}       (k_{4})
			     \frac{1}{(k_{1}+k_{2})^{2}} \biggr \},     \cr
	P^{-}_{2}= & \frac{g^{2}}{8\pi^{3}}\sum_{a=1}^{N^{2}-1}
        \sum_{i,j,k,l}^{N}(T^{a})_{ij}(T^{a})_{kl}
	\int_{0}^{\infty}\prod_{m=1}^{4}
	\frac{dk_{m}}{\sqrt{k_{m}}}
	\frac{\delta(k_{1}+k_{2}+k_{3}-k_{4})}{(k_{1}-k_{4})^2}         \cr
	& \times [{b_{i}}^{\dg}(k_{1}){b_{k}}^{\dg}(k_{2})
	        {d_{l}}^{\dg}(k_{3}) b_{j}       (k_{4})
	       +{b_{i}}^{\dg}(k_{4}) d_{k}       (k_{3})
	         b_{l}       (k_{2}) b_{j}       (k_{1})                \cr
	&      +{d_{j}}^{\dg}(k_{1}){d_{l}}^{\dg}(k_{2})
	        {b_{k}}^{\dg}(k_{3}) d_{i}       (k_{4})
	       +{d_{j}}^{\dg}(k_{4}) b_{l}       (k_{3})
	         d_{k}       (k_{2}) d_{i}       (k_{1})],              \cr
        }$$
where the integration stands for the Cauchy's principal-value
one. The Hamiltonian does not involve any term having
the creation operators only or the annihilation ones only. This
indicates that the Fock vacuum is an eigenstate of $P^-$,
i.e. the true vacuum. The property of the Hamiltonian
stems from the conservation
of the total light-cone momentum.
Each particle must have either zero
or a positive momentum, as shown in Eq. (2.8). The creation or
the annihilation
of particles, each with positive $k^{+}$, breaks the conservation.
An exception is the zero mode ($k^+=0$): Only the mode can make
the true vacuum non-trivial without breaking the conservation.
The mode is thus responsible for
non-trivial structure of vacua such as spontaneous symmetry breaking.
In the present model, however,
the mode is prohibited as long as $m \ne 0 $, because the mass term
in $P_{\rm free}^-$ enforces the eigenstate of $P^-$
to vanish at $k^+=0$ [20].

      There appears a force,
${b_{i}}^{\dg}(k_{1}){d_{i}}^{\dg}(k_{2})b_{j}(k_{3})d_{j}(k_{4})$,
in $P^{-}_{0}$, after the summation is made over $a$.
The force is considered to be induced by the so-called
annihilation diagrams where
a ${\rm q}$-$\bar {\rm q}$ pair annihilates into a
instantaneous gluon at a vertex while another pair is created
at the second vertex.  Further discussion will be made in sec. 4.

\vskip 2.0 em
\leftline{2.2. Hadronic color-singlet states}
\vskip 1.0 em

     The conserved color charges $Q^a$ ( $a=1 \sim N^2-1$ ) are
generators of SU($N$). These can be recombined into $N-1$
operators being mutually commutable and $N(N-1)/2$ pairs of raising and
lowering operators. Whenever these operators act on
color-singlet states, the value is always zero. Using the condition,
one can easily construct color-singlet states of meson and baryon,
	$$
	|\Psi_{\rm meson} \rangle =|{\rm meson}\rangle_{2} +
    |{\rm meson}\rangle_{4},                                  \eqno{\rm (2.15)}
	$$
	$$
	|{\rm meson}\rangle_{2} = \frac{1}{\sqrt{N}}\int_{0}^{\cal P}
	         \frac{dk_{1}dk_{2}}{2\pi\sqrt{k_{1}k_{2}}}
	         \delta({\cal P}-k_{1}-k_{2})\psi_{2}(k_{1},k_{2})
		 \sum_{m=1}^{N} {b_{m}}^{\dg}(k_{1}){d_{m}}^{\dg}(k_{2})
		 |0\rangle,                                   \eqno{\rm (2.16)}
	$$
	$$\eqalign{
	|{\rm meson}\rangle_{4} =\frac{1}{\sqrt{2N(N+1)}}\int_{0}^{\cal P}
	        \prod_{i=1}^{4} & \frac{dk_{i}}{\sqrt{2\pi k_{i}}}
		 \delta({\cal P}-\sum_{i=1}^{4}k_{i})  \cr
	\times \biggl\{ \psi_{\rm A}(k_{1},k_{2},k_{3},k_{4}) & \sum_{m=1}^{N}
	        {b_{m}}^{\dg}(k_{1}){d_{m}}^{\dg}(k_{2})
		{b_{m}}^{\dg}(k_{3}){d_{m}}^{\dg}(k_{4})   \cr
        +\biggl[ \psi_{\rm A}(k_{1},k_{2},k_{3},k_{4})
	        + & \sqrt{\frac{N+1}{N-1}}
		\psi_{\rm S}(k_{1},k_{2},k_{3},k_{4}) \biggr]  \cr
	\times & \sum_{m\not= n}^{N}
	        {b_{m}}^{\dg}(k_{1}){d_{m}}^{\dg}(k_{2})
		{b_{n}}^{\dg}(k_{3}){d_{n}}^{\dg}(k_{4})  \biggr\}
	|0\rangle,                                         \cr}
     \eqno{\rm (2.17)}
	$$
	$$
	|\Psi_{\rm baryon} \rangle =
	\int_{0}^{\cal P}\delta({\cal P}-\sum_{i=1}^{4}k_{i})
	\psi_{\rm b}(k_{1},k_{2},\cdots,k_{N})
        \prod_{i=1}^{N}\frac{dk_{i}}{\sqrt{2\pi k_{i}}}
	        {b_{1}}^{\dg}(k_{1})\cdots{b_{N}}^{\dg}(k_{N})
	|0\rangle,                                            \eqno{\rm (2.18)}
	$$
where the wave functions have some symmetries
for the interchange of two momenta,
	$$
	 \psi_{\rm A}(k_{1},k_{2},k_{3},k_{4})
       =-\psi_{\rm A}(k_{3},k_{2},k_{1},k_{4})
       =-\psi_{\rm A}(k_{1},k_{4},k_{3},k_{2}),               \eqno{\rm (2.19)}
	$$
	$$
	 \psi_{\rm S}(k_{1},k_{2},k_{3},k_{4})
       =\ \ \psi_{\rm S}(k_{3},k_{2},k_{1},k_{4})
       =\ \ \psi_{\rm S}(k_{1},k_{4},k_{3},k_{2}),            \eqno{\rm (2.20)}
	$$
	$$
	\psi_{\rm b}(\cdots,k_{i},\cdots,
	k_{j},\cdots)
       =\psi_{\rm b}(\cdots,k_{j},\cdots,
	k_{i},\cdots)                                        \eqno{\rm (2.21)}
	$$
for any $i$ and $j$. The color-singlet states are expanded in terms
of the number of quarks and antiquarks, and truncated
to the 2- and 4-body components in the case of the mesonic state
and to the $N$-body
one in the case of the baryonic state. The $Q^a$'s do not couple
the truncated
space with the remainder, so they keep proper commutation relations
between them within the truncated space. The truncation, i.e.
the Tamm-Dancoff approximation [10],
thus does not break the SU($N$) symmetry.
This is an advantage of LFTD.
In the equal-time quantization, the corresponding charge
operators contain terms having $b_{i}d_{j}$ or $b_{i}^{\dg}d_{j}^{\dg}$
which combine the truncated space with the remainder.

\vskip 2.0 em
\leftline{2.3.  Light-front Einstein-Schr\"odinger (ES)
equation for hadronic mass}
\vskip 1.0 em

     The ES equation for hadronic mass is $2{\cal P}P^-|\Psi \rangle =
M^2|\Psi \rangle $ in the light-front form, where $P^+$ has been replaced
by its eigenvalue ${\cal P}$ as a constant of motion. In the equation,
${\cal P}$ can be scaled out by changing variables $k_i$ into
their fractions $x_i=k_i/{\cal P}$:
The 2- and 4-body wave functions,
$\psi_2(k_1,k_2)$ and $\psi_4(k_1,k_2,k_3,k_4)$, are
also replaced by $\psi_2(x_1,x_2)$ and $\psi_4(x_1,x_2,x_3,x_4)/{\cal P}$
in $|\Psi_{\rm meson } \rangle$,
while $\psi_{\rm b}(k_1,\cdots,k_N)$ by
$\psi_{\rm b}(x_1,\cdots,x_N)/{\cal P}^{N/2-1}$
in $|\Psi_{\rm baryon}\rangle$.
Left-multiplying the rescaled equation
by individual basis of the truncated Fock space leads to a set of coupled
equations for the wave functions. In the mesonic case, the equations
are lengthy and then presented in Appendix A.
In the equations, couplings between
the 2- and 4-body sectors are of order $1/\sqrt{N}$ in the $1/N$ expansion,
where $g^2N$ is fixed.
In the large $N$ limit, the 2-body sector is then decoupled from
the 4-body one, and the 2-body sector tends to the 't Hooft
equation. ( This discussion is even clearer in the matrix representation
of the coupled equations in Appendix C. )
This conclusion is not changed by further inclusion of
the 6-body states, since the resultant equations involve
no direct coupling between the 2- and 6-body sectors. According to
numerical calculations done in sec. 3, the decoupled 4-body sector
does not produce any bound state. All the bound states in the large
$N$ limit thus appear as two-body states.

    To see the behavior of the mesonic mass in the massless limit
($m/g \rightarrow 0$), we integrate Eq. (A.1) over $x$,
$$
     M^2\int_{0}^{1}dx\psi_2(x,1-x)
    =m^2\int_{0}^{1}dx\bigg(\frac{1}{x}+\frac{1}{1-x}\bigg)
                   \psi_2(x,1-x), \eqno{\rm (2.22)}
$$
where all interaction terms have been completely canceled to each other.
The equation shows that $M=0$ and/or $\int \psi_2 dx =0$ at $m=0$.
The first condition says that $M=0$ for the ground state, and
the second one that all excited states giving positive $M$
are orthogonal to 1. Only a state orthogonal to all excited states
is the ground state, so $\psi_2=1$ for the ground state.
When $\psi_2=1$, all couplings between the 2- and 4-body sectors
vanish, so that $\psi_{\rm A}=\psi_{\rm S}=0$. It turns out that
$M=0$ and $\psi_2=1$ and $\psi_{\rm A}=\psi_{\rm S}=0$ for the ground state.
This fact suggests that the ground state has
small 4-body components in the region of $m^2 \ll g^2N$.
This will be supported by numerical tests in sec. 3.

     The equation for the baryonic wave function $\psi_{\rm b}$ is

$$\eqalign{
        & M^2\psi_{\rm b}(x_{1},x_{2},\cdots,x_{N})                     \cr
       = & \biggl(m^{2}-\frac{N^{2}-1}{2N}\frac{g^{2}}{\pi} \biggr)
	\biggl(\sum_{i=1}^{N} \frac{1}{x_{i}} \biggr)
        \psi_{\rm b}(x_{1},x_{2},\cdots,x_{N}) \cr
      & - \frac{N+1}{2N}\frac{g^{2}}{\pi}\int_{0}^{1}dy_{1}dy_{2}
	\sum_{i>j}^{N}
	\frac{\delta(x_{i}+x_{j}-y_{1}-y_{2})}{(x_{i}-y_{1})^{2}}
	\psi_{\rm b}(x_1,\cdots,y_{1},\cdots,y_{2},
	\cdots,x_N),   \cr}                                   \eqno{\rm (2.23)}
$$
where $\sum_{i=1}^Nx_i=1$ and the $i$-th and $j$-th arguments
of $\psi_{\rm b}$ in the second term have
been replaced by $y_{1}$ and $y_{2}$, respectively.
Again, the equation reduces to Eq. (2.22), when it is integrated over
all $x_i$. Equation (2.22) is still derivable,
even if the truncated space is
extended up to the $(N+2)$-body state $\psi_{N+2}$. This is explicitly
shown in Appendix A for the case of SU(2). Hence, the baryonic mass
as well as the mesonic one vanishes in the massless limit,
as far as the ground state is concerned. The baryonic wave function is
then $\psi_{\rm b}=1$ and $\psi_{N+2}=0$. Just like the mesonic case,
this implies that $\psi_{N+2}$ nearly equals to $0$
in the region of $m^2 \ll g^2N$.
For this reason, the $(N+2)$-body component will be neglected
in numerical calculations done in sec. 3.

\vskip 2.0 em
\leftline{2.4.  Approximate solutions to the lightest hadron masses}
\vskip 1.0 em

     The 2-body sector of the coupled equations for mesonic wave functions
agrees with the corresponding one
in the massive Schwinger model, except for the factor $N(1-1/N^2)$
irrelevant to the following statement. 't Hooft [1] and Bergknoff [20]
showed in the massive Schwinger model
that the term in $P^-$ involving $m$ enforces
$\psi_2(0)=\psi_2(1)=0$. Following their analysis, we can determine
the behavior of $\psi_2$ near $x=0$ and 1 as $[x(1-x)]^{\beta}$ with
$$
      2\pi m^2/[g^2N(1-1/N^2)]-1 + \pi\beta\cot (\pi\beta)=0, \eqno{\rm (2.24)}
$$
where it is assumed that $2\pi m^2/[g^2N(1-1/N^2)] \ll 1$.
In the massless limit $[x(1-x)]^{\beta}$ tends to 1, that is,
the exact solution at $m=0$, because of $\beta = 0 $ there.
This strongly implies that $[x(1-x)]^{\beta}$ is a good approximation to
$\psi_2(x)$ at all $x$, as long as $2\pi m^2/[g^2N(1-1/N^2)] \ll 1$.
This will be supported by numerical tests in sec. 3.
The same discussion can be made for baryon. Inserting
$\psi_2(x)=[x(1-x)]^{\beta}$ or $\psi_{\rm b}=[x_1x_2 \cdots x_N]^{\beta}$
into Eq.(2.22), one can obtain $M$ of the lightest state
in the region of $g^2N(1-1/N^2) \gg m^2$ as
$$
      \sqrt{2Cm \sqrt{1-\frac{1}{N^2}}}                       \eqno{\rm (2.25)}
$$
for the mesonic case and
$$
      \sqrt{CmN(N-1)\sqrt{1-\frac{1}{N^2}}}                   \eqno{\rm (2.26)}
$$
for the baryonic case, where $C=(g^2N\pi/6)^{1/2}$. The approximate
solutions show $m$- and $N$-dependences of $M$ explicitly.

     In the $1/N$ expansion of the solutions, the leading order is
$O(N^0)$ for the mesonic mass and $O(N)$ for the baryonic mass,
as expected from topological considerations [15$-$17].
As an interesting result,
the next-to-leading order is not $O(N^{-1})$ but $O(N^{-2})$
for the mesonic mass.  This makes the $1/N$
expansion more reliable especially for the lightest mesonic mass.
This is not the case for other mesonic states and the lightest
baryonic one.
     The approximate solutions also indicate that the hadronic masses behave
like $m^{0.5}$ for small $m$. This behavior is also seen in the massive
Schwinger model with two flavors[14]. As a result of the behavior, the "pion"
decay constant becomes really a constant,
indicating that PCAC is a valid
concept even for the toy model. This is true also for the present model.

\vskip 2.0 em
\leftline{\bf 3. NUMERICAL METHOD AND RESULTS}
\vskip 1.0 em
\leftline{3.1. Basis functions}
\vskip 1.0 em

%%\eqnostart = (3.1)

     The truncated ES equations for hadron masses are numerically
solved with the variational method: The wave functions are expanded in
terms of basis functions, and the coefficients of the expansion are
determined by diagonalizing $P^-$ in the space spanned by the basis
functions. All tools needed for computations are shown in Ref.
[14].

     A reasonable choice of the basis functions is
$$
     \psi_2(x,1-x)=\sum_{n=0}^{N_2} a_n f_n(x),               \eqno{\rm (3.1)}
$$
$$
     \psi_{\rm A}(x_1,x_2,x_3,x_4)
     =\sum_{n=0}^{N_4} b_n G_n(x_1,x_2,x_3,x_4),
                                                               \eqno{\rm (3.2)}
$$
$$
     \psi_{\rm S}(x_1,x_2,x_3,x_4)
     =\sum_{n=0}^{N_4} c_n G_n(x_1,x_2,x_3,x_4),
                                                               \eqno{\rm (3.3)}
$$
$$
     \psi_{\rm b}(x_1,\cdots,x_N)=\sum_{n=0}^{N_b} d_n F_n(x_1,\cdots,x_N)
                                                               \eqno{\rm (3.4)}
$$
with
$$
     f_n=\cases{ [x(1-x)]^{\beta+n}\cr [x(1-x)]^{\beta+n}(2x-1)\cr },
                                                               \eqno{\rm (3.5)}
$$
$$
     G_n=\cases{
        (x_1x_2x_3x_4)^{\beta}
        (x^-_{13})^{n_1}(x^+_{13}x^+_{24})^{n_2}(x^-_{24})^{n_3} \cr
        (x_1x_2x_3x_4)^{\beta}
        (x^-_{13})^{n_1}(x^+_{13}x^+_{24})^{n_2}
        (x^-_{24})^{n_3}x^+_{13} \cr}
                                                               \eqno{\rm (3.6)}
$$
for all $N$ and
$$
     F_n(x,1-x)=[x(1-x)]^{\beta+n}                             \eqno{\rm (3.7)}
$$
for SU(2) and
$$
     F_n(x_1,x_2,x_3)=\cases{
          (x_1x_2x_3)^{\beta}{\cal S}
          [(x^-_{12})^{n_1}(x^+_{12})^{n_1}x_{3}^{n_2}] \cr
          (x_1x_2x_3)^{\beta}{\cal S}
          [(x^-_{12})^{n_1}(x^+_{12})^{n_1}x_{3}^{n_2+1}] \cr}
      \eqno{\rm (3.8)}
$$
for SU(3), where $\sum x_i=1$ for each basis function,
$x^{\pm}_{ij}=x_i \pm x_j$ and ${\cal S}$ is the symmertrizer.
The subscript $n$ of $G_n$ stands for a set $(n_1,n_2,n_3)$,
and for other functions analogously.
As already discussed in sec. 2.4, $f_0$ ( $F_0$) is a good
approximation to the exact $\psi_2$ ( $\psi_{\rm b}$ ).
Each type of basis functions forms
a complete set, when the upper limit
$N_n ( n=2, 4,{\rm b})$ of the summation is infinite.
The $G_n$'s are constructed from the set
$ \{ (x_1x_2x_3x_4)^{\beta}x_1^{n_1}x_2^{n_2}x_3^{n_3}x_4^{x_4} \} $
which obviously form a complete set. First, it is transformed into
$ \{ (x_1x_2x_3x_4)^{\beta}
(x^-_{13})^{n_1}(x^+_{13})^{n_2}(x^+_{24})^{n_3}(x^-_{24})^{n_4} \} $.
Next, the factor $(x^+_{13})^{n_2}(x^+_{24})^{n_3}$ in the set
is expanded in terms of $ ( x^+_{13}x^+_{24} ) ^n$ and
$ (x^+_{13}x^+_{24})^nx^+_{13}$,
where $x^+_{13}+x^+_{24}=1$. ( See Appendix B.)  The final form is
Eq.(3.6),
in which the number of summations has been reduced from 4 to 3.
This is a merit of this form. Another merit is that the symmetry
for the interchange of $x_1$ and $x_3$
( $x_2$ and $x_4$ ) is easily imposed on $ \psi_4 $
by taking either even or odd $n_1$ ( $n_3$ ).
Similar consideration is made for $f_n$ and $F_n$.

\vskip 2.0 em
\leftline{3.2.  Numerical results}
\vskip 1.0 em

     In general, $M$ calculated with the variational method
depends on $N_{\alpha}$  ($\alpha=2,4,{\rm b}$)
which characterizes a size of
the space spanned by the basis functions, unless the space is large enough
to yield an accurate $M$. In the present calculation,
the space would be sufficiently large, since the dependence is very weak,
owing to the effective choice of basis functions. This is shown
in Fig. 1 for the case of SU(2) meson. Fig. 1(a) represents
the $N_2$-dependence of the lightest mass ($M_1$) and
the second lightest one ($M_2$), while Fig. 1(b) does their $N_4$-dependence.
Hereafter, $m$ and $M$ are presented in units of $\sqrt{g^2N/2\pi}$.
In the case of $m=10^{-4}$,
$M_1$ and $M_2$ converge at $(N_2, N_4)=(4,4)$. For the baryonic mass
($M_{\rm b}$),
convergence is seen at $N_{\rm b}=2$.
Our full-fledged calculations are then done with $(N_2, N_4)=(4,4)$
for the mesonic case and with $N_{\rm b}=2$ for the baryonic one.

     The $m$-dependence of hadron masses obtained
with full-fledged calculations is shown in Table 1 and Fig. 2
for both SU(2) and SU(3).
There are two mesonic and one baryonic bound states
in the range $m < 0.1$. The hadronic masses
behave like $m^{0.5}$ as $m \rightarrow 0$,
as predicted by the approximate solutions ( Eqs. (2.25) and (2.26))
to $M_1$ and $M_{\rm b}$.
Ratios $M_1/2M_{\rm b}$ and $M_2/2M_{\rm b}$
at small $m$, say $m=10^{-4}$,
are 0.4585 and 0.8122 for SU(2) and 0.2886 and 0.5722 for SU(3),
while the corresponding
results, $\sin\{\frac{\pi n}{2(2N-1)}\}$ ($n=1,2$),
of the bosonization in the massless limit
are 0.5000 and 0.8660 for SU(2) and 0.3090 and 0.5877 for SU(3).
The two types of results are identical
within error of $\sim10 \%$ for SU(2) and of $\sim5 \%$ for SU(3).
In general, our calculations of the baryonic mass are
relatively inaccurate compared with those of the mesonic masses,
since the truncated Fock space
is smaller in the baryonic case than in the mesonic one; to be precise,
only the valence ($N$-body) state is included in the baryonic Fock space,
while both the valence (q$\bar{\rm q}$) and the 4-body
(qq$\bar{\rm q}\bar{\rm q}$) state are included in the mesonic space.
In the SU(2) case, $M_1$ is reduced by $\sim 10\%$ by extending the
Fock space from the 2-body subspace to the 2-body plus 4-body
one.  It is very likely that such a reduction takes place also for
$M_{\rm b}$, since the ES equation
for $M_{\rm b}$ is very similar to that for $M_1$ in the SU(2) case,
as shown in Appendix A.
The $\sim 10 \%$ error for SU(2) thus might come from the fact that the
$(N+2)$-body state is not included in the baryonic Fock space.
In the SU(3) case, on the other hand, $M_1$ is not changed
by the extension. It is then likely that
$M_{\rm b}$ is also unchanged by a similar extension of the Fock space
from the $N$-body subspace to the $N$-body plus $(N+2)$-body one.
Hence, our calculations might be accurate for the SU(3) baryonic mass.
An unsettled problem is what causes the $\sim5 \%$ error for SU(3).
This will be discussed in sec. 4.

     Unfortunately, our results can not be compared with
those of DLCQ in Ref. [6] ; hadronic masses presented
by some figures and a table in the paper are inconsistent.

     For the lightest mesonic state of SU(2),
the probability ($P_2$) of being in the 2-body state is
much larger than that ($P_4$) in the 4-body state;
$P_2=98.3 \%$ and $P_4=1.7 \%$ at $m=10^{-4}$.
The lightest mesonic state is odd under charge conjugation,
because the 2-body component
is symmetric under $x_1 \leftrightarrow x_2$.
The second lightest state is, on the other hand, highly relativistic
in the sense that
$P_2 \sim P_4$; $P_2=42.9 \%$ and $P_4=57.1 \%$ at $m=10^{-4}$ for SU(2).
This state is even under charge conjugation,
because the 2-body piece
is antisymmetric under $x_1 \leftrightarrow x_2$.

     The approximate solutions
(Eqs.(2.25) and (2.26)) to $M_1$ and $M_{\rm b}$
are compared with
numerical ones obtained with the full-fledged calculations, in two cases of
SU(2) and SU(3). For SU(2), the approximate $M_1$ agrees with
the approximate $M_{\rm b}$, as shown in Eqs. (2.25) and (2.26).
They are depicted by the common dashed line in Fig. 3(a),
and compared with the numerical solutions for $M_1$ (solid line) and
for $M_{\rm b}$ (dot-dashed line).
The numerical and approximate solutions consist with each other
at $m < 0.001$ for $M_1$ and at $m < 0.03$ for $M_{\rm b}$.
For SU(3) in Fig. 3(b),
the approximate solutions well reproduce the numerical ones,
for both $M_1$ and $M_{\rm b}$, at $m < 0.1$.
The agreement would be seen also
at $N$ larger than 3; this is true at least for $M_1$ (see Fig. 4).
The $N$-dependence of $M_1$ and $M_{\rm b}$
is thus obtained accurately with the
approximate solutions, as long as $m^2 \ll 1$.

    The $N$-dependence of $M_1$ and $M_2$ is shown
in Fig. 4(a) for the case of
$m=10^{-4}$. $N$ is varied widely from 2 to $\infty$.
The approximate solution to $M_1$ (dashed line) well simulates
the numerical solution (solid line). As expected from
the weak $N$-dependence of the approximate $M_1$, $M_1$ at small $N$
is close to that at $N\to\infty$ (the lightest 't Hooft mass ).
The second mass is below the threshold ($2M_1$) for $N=2$ and 3, but
not for $N\geq4$. The second mesonic state is thus bound only for small $N$
such as 2 and 3. For $N\to\infty$, on the other hand, it
becomes a 4-body unbound state absent in
the 't Hooft solution. The $1/N$ expansion thus
works well for the first mesonic mass, but not for the second mass.

     The $N$-dependence of $M_{\rm b}$ is shown in Fig. 4(b).
The approximate solution to $M_{\rm b}$ ( the solid line )
well reproduces the numerical result ( the dashed line) for $N=2$ and 3.
If the mass is expanded in $1/N$ and truncated to the leading,
as in Ref. [15$-$17],
the mass should be proportional to $N$. The ratio of $M_{\rm b}$
at $N=3$ to that at $N=2$ is, however, 1.807 and larger than 3/2.
This indicates that the higher-order terms are not negligible for such
small $N$.

     Figure 5 shows the $m$-dependence of mesonic masses
in the large $N$ limit. In this limit, the 2- and 4-body sectors of
the ES equations are decoupled with each other, so that all states
appear as either 2- or 4-body state. Obviously, all the 2-body states
( the 't Hooft solutions ) are bound. In Fig. 5, on the other hand,
the lightest 4-body state is above the threshold ($2M_1$),
where $M_1$ is the mass of the lightest 2-body bound state.
All the 4-body states are thus unbound. This is understandable from the
statement [13] based on the $1/N$ expansion
that two mesons ( two ${\rm q}$-$\bar {\rm q}$ pairs ) do not interact
in the limit.

\vskip 2.0 em
\leftline{\bf 4. DISCUSSIONS}
\vskip 1.0 em

     Three unsettled problems are discussed.

(1) As shown in sec. 3.2, our results for
$M_1/2M_{\rm b}$ and $M_2/2M_{\rm b}$ deviate from the predictions
of the bosonization [2] by $\sim 5\%$ in the case of SU(3).
A problem in the comparison is
that the two types of results are obtained at different $m$;
the results of the bosonization at $m=0$ and ours at $m=10^{-4}$.
In the bosonization, the color-singlet
bosonic field is coupled with the non-singlet fields.
The non-singlet fields are, however, neglected in the bosonic form
of the Hamiltonian. The neglect seems to induce an error of
$O(M_{\rm S}/M_{\rm NS})$, where $M_{\rm S}$ ($M_{\rm NS}$) is a mass
of the states generated by the singlet (non-singlet) field.
We intuitively
think that the non-singlet fields can generate singlet states
as a result of their superposition,
but it is not clear from the bosonic form of
the Hamiltonian which does not possess the SU($N$) symmetry explicitly.
The $M_{\rm NS}$ seems to be $O(1)$ independently of $m/g$,
since the non-singlet fields
have mass terms of $O(1)$. The $M_{\rm S}$, on the other hand,
tends to zero in the massless limit ($m/g \rightarrow 0$).
The result of the bosonization is then correct in the massless limit.
At $m =10^{-4}$, it has an error of
$O(M_{\rm S}/M_{\rm NS})=O(M_{1}/M_{\rm NS})=10^{-2}$.
This error is a possible origin of
the discrepancy between the two types of results.

(2) Our truncated Fock space consists of the 2- and 4-body states
in the mesonic case. Further inclusion of 6-body states would
produce the third mesonic bound state in the strong coupling region,
because the bosonization [2] predicts for SU($N$) meson
that there appear $2N-1$ massless bound states
in the strong coupling limit, and because DLCQ [6] concludes
that there are many massless mesonic states in the limit and
the $n$-th state consists of $n$ components
from 2-body to $2n$-body. The conclusion of DLCQ also
implies that the second mesonic state could be described accurately
within the present truncated Fock space, as long as $g^2N/m^2 \gg 1$.

(3) In the Schwinger model, the axial symmetry is anomalous and
the light-cone Hamiltonian involves a force induced by
the so-called annihilation diagrams where
a ${\rm q}$-$\bar {\rm q}$ pair annihilates into a
instantaneous gluon at a vertex while another pair is created
at the second vertex [14].
The force generates a term $2\int_0^1 dx \psi(x,1-x)$
in the ES equation for "$\eta$" (iso-singlet) mass, but not for
"$\pi$" (iso-triplets) mass.
The term splits the $\eta$ mass from the $\pi$ mass; especially
in the chiral limit, the $\eta$ mass keeps a finite value, but
the $\pi$ mass vanishes.
In the present model, on the other hand,
the axial symmetry is not anomalous, but the annihilation force is
still in $P^-$. In this case, the force does not
generate any term which makes $\eta$ massive in the chiral limit.
Thus, the $\eta$-$\pi$ splitting
( the U(1) problem ) may not be resolved simply as a matter of
the annihilation force.

      Throughout this work, we conclude that
LFTD is a powerful tool for computing non-perturbative quantities
such as hadronic masses. LFTD might be more useful
than the $1/N$ expansion and the bosonization
which are valid only in a particular situation such as the large $N$ and
$g/m$ limits.
\vfill\eject
%
%\vskip 2.0 em
\leftline{\bf ACKNOWLEDGMENTS}
\vskip 1.0 em

We would like to acknowledge stimulating conversations with our
colleagues, in particular, Dr. K. Harada.
\vskip 2 true cm
%%eqnostart = (A.1)
%
%
%
%
%
%
%
\leftline{\bf Appendix A: EINSTEIN-SCHR\"ODINGER EQUATIONS}
   A set of coupled integral equations is
obtained by applying the Hamiltonian (2.14) to the states
(2.15) $\sim$ (2.18). For the 4-body wave
functions, $\sum_{i=1}^4x_i=1$. It reads, for SU($N$) mesons,
$$\eqalign{
        & M^2\psi_{2}(x,1-x)                                    \cr
       = & \biggl(m^{2}-\frac{N^{2}-1}{2N}\frac{g^{2}}{\pi} \biggr)
	\biggl(\frac{1}{x}+\frac{1}{1-x} \biggr)\psi_{2}(x,1-x) \cr
       & -\frac{N^{2}-1}{2N}\frac{g^{2}}{\pi}
	\int_{0}^{1} dy\frac{\psi_{2}(y,1-y)}{(x-y)^{2}}        \cr
       & +\frac{(N-1)\sqrt{2(N+1)}}{2N}\frac{g^{2}}{\pi}
	\int_{0}^{1}dy_{1}dy_{2}dy_{3}                          \cr
       &\hskip 2.1 true cm
	\times \ \biggl\{ \psi_{\rm A}(x,y_{1},y_{2},y_{3})
	\frac{\delta(y_{1}+y_{2}+y_{3}-(1-x))}{(y_{3}-(1-x))^{2}} \cr
       &\hskip 2.5 true cm-\psi_{\rm A}(y_{1},y_{2},y_{3},1-x)
	\frac{\delta(y_{1}+y_{2}+y_{3}-x)}{(y_{1}-x)^2} \biggr\}  \cr
       & +\frac{(N+1)\sqrt{2(N-1)}}{2N}\frac{g^{2}}{\pi}
	\int_{0}^{1}dy_{1}dy_{2}dy_{3}                          \cr
       &\hskip 2.1 true cm
	\times \ \biggl\{ \psi_{\rm S}(x,y_{1},y_{2},y_{3})
	\frac{\delta(y_{1}+y_{2}+y_{3}-(1-x))}{(y_{3}-(1-x))^{2}}\cr
       &\hskip 2.5 true cm -\psi_{\rm S}(y_{1},y_{2},y_{3},1-x)
	\frac{\delta(y_{1}+y_{2}+y_{3}-x)}{(y_{1}-x)^2} \biggr\}, \cr}
                                                            \eqno{\rm (A.1)}
$$
\vfill\eject
%
%\vskip 1.0 true cm
%
$$\eqalign{
        & M^2\psi_{\rm A}(x_{1},x_{2},x_{3},x_{4})                  \cr
       = & \biggl(m^{2}-\frac{N^{2}-1}{2N}\frac{g^{2}}{\pi} \biggr)
	\biggl(\sum_{i=1}^{4} \frac{1}{x_{i}} \biggr)
        \psi_{\rm A}(x_{1},x_{2},x_{3},x_{4})                       \cr
       & +\frac{(N-1)\sqrt{2(N+1)}}{8N}\frac{g^{2}}{\pi}        \cr
& \times
\biggl\{
        \biggl[\frac{1}{(x_{2}+x_{3})^{2}}
              -\frac{1}{(x_{3}+x_{4})^{2}}\biggr]\psi_{2}(x_{1},1-x_{1}) \cr
&\hskip 0.3 true cm  + \biggl[\frac{1}{(x_{3}+x_{4})^{2}}
              -\frac{1}{(x_{1}+x_{4})^{2}}\biggr]\psi_{2}(1-x_{2},x_{2}) \cr
&\hskip 0.3 true cm  + \biggl[\frac{1}{(x_{1}+x_{4})^{2}}
              -\frac{1}{(x_{1}+x_{2})^{2}}\biggr]\psi_{2}(x_{3},1-x_{3}) \cr
&\hskip 0.3 true cm  + \biggl[\frac{1}{(x_{1}+x_{2})^{2}}
              -\frac{1}{(x_{2}+x_{3})^{2}}\biggr]\psi_{2}(1-x_{4},x_{4})
\biggr\}   \hskip 3.0 true cm     \cr
    & + \frac{N-1}{2N}\frac{g^{2}}{\pi}\int_{0}^{1}dy_{1}dy_{2}
\biggl\{
	\frac{\delta(x_{1}+x_{3}-y_{1}-y_{2})}{(x_{1}-y_{1})^{2}}
	\psi_{\rm A}(y_{1},x_{2},y_{2},x_{4})  \cr
&\hskip 4 true cm
	+ \frac{\delta(x_{2}+x_{4}-y_{1}-y_{2})}{(x_{2}-y_{1})^{2}}
	\psi_{\rm A}(x_{1},y_{1},x_{3},y_{2})
\biggr\}   \cr
        & \quad \cr
    & + \frac{N-1}{4N}\frac{g^{2}}{\pi}\int_{0}^{1}dy_{1}dy_{2}
\biggl\{
	\frac{\delta(x_{1}+x_{2}-y_{1}-y_{2})}{(x_{1}+x_{2})^{2}}
	\psi_{\rm A}(y_{1},y_{2},x_{3},x_{4})  \cr
&\hskip 4 true cm
	- \frac{\delta(x_{1}+x_{4}-y_{1}-y_{2})}{(x_{1}+x_{4})^{2}}
	\psi_{\rm A}(y_{1},y_{2},x_{3},x_{2})  \cr
&\hskip 4 true cm
	- \frac{\delta(x_{3}+x_{2}-y_{1}-y_{2})}{(x_{3}+x_{2})^{2}}
	\psi_{\rm A}(y_{1},y_{2},x_{1},x_{4})  \cr
&\hskip 4 true cm
	+ \frac{\delta(x_{3}+x_{4}-y_{1}-y_{2})}{(x_{3}+x_{4})^{2}}
	\psi_{\rm A}(y_{1},y_{2},x_{1},x_{2})
\biggr\}  \cr
        & \quad \cr
    & - \frac{(N-1)(N+2)}{4N}\frac{g^{2}}{\pi}\int_{0}^{1}dy_{1}dy_{2}
\biggl\{
	\frac{\delta(x_{1}+x_{2}-y_{1}-y_{2})}{(x_{1}-y_{1})^{2}}
	\psi_{\rm A}(y_{1},y_{2},x_{3},x_{4})  \cr
&\hskip 5.7 true cm
	- \frac{\delta(x_{1}+x_{4}-y_{1}-y_{2})}{(x_{1}-y_{1})^{2}}
	\psi_{\rm A}(y_{1},y_{2},x_{3},x_{2})  \cr
&\hskip 5.7 true cm
	- \frac{\delta(x_{3}+x_{2}-y_{1}-y_{2})}{(x_{3}-y_{1})^{2}}
	\psi_{\rm A}(y_{1},y_{2},x_{1},x_{4})  \cr
&\hskip 5.7 true cm
	+ \frac{\delta(x_{3}+x_{4}-y_{1}-y_{2})}{(x_{3}-y_{1})^{2}}
	\psi_{\rm A}(y_{1},y_{2},x_{1},x_{2})  \bigg\}  \cr }
$$
\vfill\eject
$$\eqalign{
        & \quad \cr
    & - \frac{\sqrt{(N+1)(N-1)}}{4N}\frac{g^{2}}{\pi}\int_{0}^{1}dy_{1}dy_{2}
\cr
&\hskip 1.5 true cm \times
\biggl\{
	\delta(x_{1}+x_{2}-y_{1}-y_{2})\psi_{\rm S}(y_{1},y_{2},x_{3},x_{4})
\biggl [\frac{1}{(x_{1}+x_{2})^{2}}+\frac{N}{(x_{1}-y_{1})^{2}} \biggr]
	\cr
&\hskip 1.8 true cm
	- \delta(x_{1}+x_{4}-y_{1}-y_{2})\psi_{\rm S}(y_{1},y_{2},x_{3},x_{2})
\biggl [\frac{1}{(x_{1}+x_{4})^{2}}+\frac{N}{(x_{1}-y_{1})^{2}} \biggr]
	\cr
&\hskip 1.8 true cm
	- \delta(x_{3}+x_{2}-y_{1}-y_{2})\psi_{\rm S}(y_{1},y_{2},x_{1},x_{4})
\biggl [\frac{1}{(x_{3}+x_{2})^{2}}+\frac{N}{(x_{3}-y_{1})^{2}} \biggr]
	\cr
&\hskip 1.8 true cm
+ \delta(x_{3}+x_{4}-y_{1}-y_{2})\psi_{\rm S}(y_{1},y_{2},x_{1},x_{2})
\biggl [\frac{1}{(x_{3}+x_{4})^{2}}+\frac{N}{(x_{3}-y_{1})^{2}} \biggr]
	\biggr\},
\cr}                \eqno{\rm (A.2)}
$$
\vskip 1.5 true cm
$$\eqalign{
        & M^2\psi_{\rm S}(x_{1},x_{2},x_{3},x_{4})                  \cr
       & =\biggl(m^{2}-\frac{N^{2}-1}{2N}\frac{g^{2}}{\pi} \biggr)
	\biggl(\sum_{i=1}^{4} \frac{1}{x_{i}} \biggr)
        \psi_{\rm S}(x_{1},x_{2},x_{3},x_{4})                       \cr
       & +\frac{(N+1)\sqrt{2(N-1)}}{8N}\frac{g^{2}}{\pi}        \cr
& \times
\biggl\{
        \biggl[\frac{1}{(x_{2}+x_{3})^{2}}
              +\frac{1}{(x_{3}+x_{4})^{2}}\biggr]\psi_{2}(x_{1},1-x_{1})
\cr
&\hskip 0.3 true cm
      - \biggl[\frac{1}{(x_{3}+x_{4})^{2}}
              +\frac{1}{(x_{1}+x_{4})^{2}}\biggr]\psi_{2}(1-x_{2},x_{2})
\cr
&\hskip 0.3 true cm
      + \biggl[\frac{1}{(x_{1}+x_{4})^{2}}
              +\frac{1}{(x_{1}+x_{2})^{2}}\biggr]\psi_{2}(x_{3},1-x_{3})
\cr
&\hskip 0.3 true cm
      - \biggl[\frac{1}{(x_{1}+x_{2})^{2}}
              +\frac{1}{(x_{2}+x_{3})^{2}}\biggr]\psi_{2}(1-x_{4},x_{4})
\biggr\} \cr
      & - \frac{N+1}{2N}\frac{g^{2}}{\pi}\int_{0}^{1}dy_{1}dy_{2}
\biggl\{
	\frac{\delta(x_{1}+x_{3}-y_{1}-y_{2})}{(x_{1}-y_{1})^{2}}
	\psi_{\rm S}(y_{1},x_{2},y_{2},x_{4})  \cr
&\hskip 4 true cm
      + \frac{\delta(x_{2}+x_{4}-y_{1}-y_{2})}{(x_{2}-y_{1})^{2}}
	\psi_{\rm S}(x_{1},y_{1},x_{3},y_{2})
\biggr\}  \cr  }
$$
\vfill\eject
$$\eqalign{
    & + \frac{N+1}{4N}\frac{g^{2}}{\pi}\int_{0}^{1}dy_{1}dy_{2}
\biggl\{
	\frac{\delta(x_{1}+x_{2}-y_{1}-y_{2})}{(x_{1}+x_{2})^{2}}
	\psi_{\rm S}(y_{1},y_{2},x_{3},x_{4})   \cr
&\hskip 4 true cm
      + \frac{\delta(x_{1}+x_{4}-y_{1}-y_{2})}{(x_{1}+x_{4})^{2}}
	\psi_{\rm S}(y_{1},y_{2},x_{3},x_{2})   \cr
&\hskip 4 true cm
      + \frac{\delta(x_{3}+x_{2}-y_{1}-y_{2})}{(x_{3}+x_{2})^{2}}
	\psi_{\rm S}(y_{1},y_{2},x_{1},x_{4})   \cr
&\hskip 4 true cm
      + \frac{\delta(x_{3}+x_{4}-y_{1}-y_{2})}{(x_{3}+x_{4})^{2}}
	\psi_{\rm S}(y_{1},y_{2},x_{1},x_{2})
\biggr\}   \cr
      & - \frac{(N+1)(N-2)}{4N}\frac{g^{2}}{\pi}\int_{0}^{1}dy_{1}dy_{2}
\biggl\{
	\frac{\delta(x_{1}+x_{2}-y_{1}-y_{2})}{(x_{1}-y_{1})^{2}}
	\psi_{\rm S}(y_{1},y_{2},x_{3},x_{4})   \cr
&\hskip 5.7 true cm
      + \frac{\delta(x_{1}+x_{4}-y_{1}-y_{2})}{(x_{1}-y_{1})^{2}}
	\psi_{\rm S}(y_{1},y_{2},x_{3},x_{2})   \cr
&\hskip 5.7 true cm
      + \frac{\delta(x_{3}+x_{2}-y_{1}-y_{2})}{(x_{3}-y_{1})^{2}}
	\psi_{\rm S}(y_{1},y_{2},x_{1},x_{4})   \cr
&\hskip 5.7 true cm
      + \frac{\delta(x_{3}+x_{4}-y_{1}-y_{2})}{(x_{3}-y_{1})^{2}}
	\psi_{\rm S}(y_{1},y_{2},x_{1},x_{2})
\biggr\}  \cr
}$$
$$\eqalign{
&     - \frac{\sqrt{(N+1)(N-1)}}{4N}
        \frac{g^{2}}{\pi}\int_{0}^{1}dy_{1}dy_{2}  \cr
&\hskip 0.3 true cm \times
\biggl\{
	\delta(x_{1}+x_{2}-y_{1}-y_{2})
	\psi_{\rm A}(y_{1},y_{2},x_{3},x_{4})
\biggl [\frac{1}{(x_{1}+x_{2})^{2}}+\frac{N}{(x_{1}-y_{1})^{2}} \biggr]
	\cr
&\hskip 0.6 true cm
      + \delta(x_{1}+x_{4}-y_{1}-y_{2})
	\psi_{\rm A}(y_{1},y_{2},x_{3},x_{2})
\biggl [\frac{1}{(x_{1}+x_{4})^{2}}+\frac{N}{(x_{1}-y_{1})^{2}} \biggr]
	\cr
&\hskip 0.6 true cm
      + \delta(x_{3}+x_{2}-y_{1}-y_{2})
	\psi_{\rm A}(y_{1},y_{2},x_{1},x_{4})
\biggl [\frac{1}{(x_{3}+x_{2})^{2}}+\frac{N}{(x_{3}-y_{1})^{2}} \biggr]
	\cr
&\hskip 0.6 true cm
      + \delta(x_{3}+x_{4}-y_{1}-y_{2})
	\psi_{\rm A}(y_{1},y_{2},x_{1},x_{2})
\biggl [\frac{1}{(x_{3}+x_{4})^{2}}+\frac{N}{(x_{3}-y_{1})^{2}} \biggr]
	\biggr\}.
\cr}                                                   \eqno{\rm (A.3)}
$$
%\hskip 0.3 true cm \times
\vskip 1 true cm
  The baryonic state is truncated up to the $(N+2)$-body component.
The $(N+2)$-body component is constructed for SU(2) as
$$
   \eqalign{
      &|{\rm baryon}\rangle_{N+2} \cr
    = & \frac{1}{2}\int_0^{\cal P}
        \prod_{i=1}^4\frac{dk_i}{\sqrt{2\pi k_i}}
        \delta({\cal P}-\sum_{i=1}^4 k_i)
        \psi_{\rm M}(k_1,k_2,k_3,k_4) \cr
      & \times
     [b_1^{\dg}(k_1) b_2^{\dg}(k_2) b_1^{\dg}(k_3) d_1^{\dg}(k_4)
     -b_2^{\dg}(k_1) b_1^{\dg}(k_2) b_2^{\dg}(k_3) d_2^{\dg}(k_4)]
      |0\rangle,\cr}  \eqno{\rm (A.4)}
$$
with the symmetry
$$
              \psi_{\rm M}(k_1,k_2,k_3,k_4)
            =-\psi_{\rm M}(k_3,k_2,k_1,k_4) , \eqno{\rm (A.5)}
$$
in addition to the $N$-body component in Eq. (2.18).
The $(N+2)$-body wave function, $\psi(k_1,k_2,k_3,k_4)$, is antisymmetric
under the interchange of $k_1$ and $k_3$, because the operator,
$b_i^{\dg}(k_1) b_j^{\dg}(k_2) b_i^{\dg}(k_3) d_j^{\dg}(k_4)$, is
antisymmetric under the interchange. The wave function can be
classified with irreducible representations of the symmetric group,
$$
        \psi(k_1,k_2,k_3,k_4)
       =\psi_{\rm s}(k_1,k_2,k_3,k_4)+\psi_{\rm a}(k_1,k_2,k_3,k_4)
                                     +\psi_{\rm M}(k_1,k_2,k_3,k_4),
                                             \eqno{\rm (A.6)}
$$
where
$$\eqalign{
       \psi_{\rm s}(k_1,k_2,k_3,k_4)&=
                  {\cal S}_{123}\psi(k_1,k_2,k_3,k_4),\cr
       \psi_{\rm a}(k_1,k_2,k_3,k_4)&=
                  {\cal A}_{123}\psi(k_1,k_2,k_3,k_4),\cr
       \psi_{\rm M}(k_1,k_2,k_3,k_4)&=
                \frac{1}{3}[2\psi(k_1,k_2,k_3,k_4)
                            -\psi(k_2,k_3,k_1,k_4)
                            -\psi(k_3,k_1,k_2,k_4)], \cr} \eqno{\rm (A.7)}
$$
where ${\cal S}_{123}$ is the symmetrizer and ${\cal A}_{123}$ is
the antisymmetrizer of momenta $k_1, k_2$ and $k_3$. Only the mixed
symmetry $\psi_{\rm M}$ can survive under the condition that
$Q^2|\Psi>=0$. The $(N+2)$-body wave function
can be constructed straightforwardly for arbitrary $N$.
The coupled equations for SU(2) baryon are then
$$\eqalign{
        M^2&\psi_{\rm b}(x,1-x)  \cr
     = \bigg(m^2&-\frac{3g^2}{4\pi}\bigg)
         \bigg(\frac{1}{x}+\frac{1}{1-x}\bigg)\psi_{\rm b} (x,1-x)
        - \frac{3g^2}{4\pi}
             \int_0^1dy\frac{\psi_{\rm b}(y,1-y)}{(x-y)^2} \cr
       + \frac{g^2}{2\pi} & \int_0^1dy_1dy_2dy_3  \cr
        \times \bigg\{
        & \delta(y_1+y_2+y_3-x)
            \psi_{\rm M}(y_1,1-x,y_2,y_3)
                 \frac{1}{(x-y_1)^2}       \cr
      -2 & \delta(y_1+y_2+y_3-x)
            \psi_{\rm M}(1-x,y_1,y_2,y_3)
           \bigg[\frac{1}{2(x-y_1)^2}+\frac{1}{(x-y_2)^2}\bigg]
        \bigg\},                                 \cr}\eqno{\rm (A.8)}
$$
\vskip 0.5 true cm
$$\eqalign{
        & M^2\psi_{\rm M}(x_{1},x_{2},x_{3},x_{4})                  \cr
       = & \biggl(m^{2}-\frac{3g^2}{4\pi} \biggr)
	\biggl(\sum_{i=1}^{4} \frac{1}{x_{i}} \biggr)
        \psi_{\rm M}(x_{1},x_{2},x_{3},x_{4})                       \cr
       & +\frac{g^2}{2\pi}
\biggl\{\frac{1}{2}
            \biggl[\frac{1}{(x_3+x_4)^2}
                  -\frac{1}{(x_1+x_4)^2}\biggr]
                       \psi_{\rm b}(x_2,1-x_2)      \cr
&\hskip 1.2 true cm
          - \biggl[\frac{1}{(x_2+x_4)^2}
                 +2\frac{1}{(x_3+x_4)^2}\biggr]
                       \psi_{\rm b}(x_1,1-x_1)      \cr
&\hskip 1.2 true cm
          + \biggl[\frac{1}{(x_2+x_4)^2}
                 +2\frac{1}{(x_1+x_4)^2}\biggr]
                       \psi_{\rm b}(x_3,1-x_3)
\biggr\}  \hskip 4.0 true cm \cr }
$$
$$\eqalign{
    & - \frac{g^2}{2\pi}\int_{0}^{1}dy_1dy_2   \cr
& \hskip 1 true cm \times \biggl\{
	\delta(x_1+x_2-y_1-y_2)
     \bigg[ \frac{1}{2(x_1-y_1)^2} + \frac{1}{(x_1-y_2)^2} \bigg]
                      \psi_{\rm M}(y_1,y_2,x_3,x_4)  \cr
&\hskip 1.3 true cm +
	\delta(x_2+x_3-y_1-y_2)
     \bigg[ \frac{1}{2(x_2-y_1)^2} + \frac{1}{(x_2-y_2)^2} \bigg]
                      \psi_{\rm M}(x_1,y_1,y_2,x_4)  \cr
&\hskip 1.0 true cm -
        \frac{1}{4} \delta(x_1+x_3-y_1-y_2)
     \bigg[ \frac{1}{2(x_1-y_1)^2} - \frac{1}{(x_1-y_2)^2} \bigg]
                      \psi_{\rm M}(y_1,x_2,y_2,x_4)
\biggr\} \hskip 1.0 true cm \cr  }
$$
$$\eqalign{
    & + \frac{g^2}{2\pi}\int_{0}^{1}dy_1dy_2   \cr
& \hskip 1 true cm \times \biggl\{
	\delta(x_1+x_4-y_1-y_2)
     \bigg[ \frac{1}{4(x_1+x_4)^2} - \frac{1}{(x_1-y_1)^2} \bigg]
                      \psi_{\rm M}(y_1,x_2,x_3,y_2)  \cr
&\hskip 1.3 true cm +
	\delta(x_3+x_4-y_1-y_2)
     \bigg[ \frac{1}{4(x_3+x_4)^2} - \frac{1}{(x_3-y_1)^2} \bigg]
                      \psi_{\rm M}(x_1,x_2,y_1,y_2)  \cr
&\hskip 1.3 true cm +
        \delta(x_2+x_4-y_1-y_2)
     \bigg[ \frac{1}{(x_2+x_4)^2} + \frac{1}{2(x_2-y_1)^2} \bigg]
                      \psi_{\rm M}(x_1,y_1,x_3,y_2)  \cr
&\hskip 1.3 true cm +
	\delta(x_1+x_4-y_1-y_2)
     \bigg[ \frac{1}{4(x_1+x_4)^2} + \frac{1}{2(x_1-y_1)^2} \bigg]
                      \psi_{\rm M}(x_2,y_1,x_3,y_2)  \cr
&\hskip 1.3 true cm +
	\delta(x_3+x_4-y_1-y_2)
     \bigg[ \frac{1}{4(x_3+x_4)^2} + \frac{1}{(x_3-y_1)^2} \bigg]
                      \psi_{\rm M}(x_1,y_1,x_2,y_2)
\biggr\}. \cr  }                            \eqno{\rm (A.9)}
$$
Again, integrating Eq. (A.8) over $x$ leads to Eq. (2.22).
%
%
%
%
%%eqnostart = (B.1)
%
%
\vskip 1.0 true cm
\leftline{\bf Appendix B: BASIS FUNCTIONS}
   The function $\{x^m(1-x)^n\}$ can be expanded in terms of
$[x(1-x)]^k$ and $[x(1-x)]^l(1-2x)$, because
$$
    \eqalign{
         &x^m(1-x)^n   \cr
       =&2^{-|m-n|}\bigg\{\sum_{i=0{\rm(even})}^{|m-n|}{}_{|m-n|}C_i
                     \sum_{j=0}^{\frac{i}{2}}{}_{\frac{i}{2}}C_j
                       (-4)^j[x(1-x)]^{{\rm min}(m,n)+j}   \cr
         & -\epsilon(m-n)
                     \sum_{i=1{\rm(odd})}^{|m-n|}{}_{|m-n|}C_i
                     \sum_{j=0}^{[\frac{i}{2}]}{}_{\frac{i}{2}}C_j
                       (-4)^j[x(1-x)]^{{\rm min}(m,n)+j}(1-2x)
              \bigg\} \cr                 }. \eqno{\rm (B.1)}
$$
\vskip 0.5 true cm
%
%%%eqnostart = (C.1)
%
\leftline{\bf Appendix C: MATRIX EIGENVALUE EQUATIONS}

      The following eigenvalue equations of matrix form are obtained
from the coupled equations (Eqs. (A.1) to (A.3)) by sandwiching them
with individual 2- and 4-body basis functions.
$$
	M^2\left(\matrix{A^{(1)}&0&0          \cr
                         0&B^{(1)}&0          \cr
                         0&0&B^{(1)}          \cr} \right)
           \left(\matrix{a  		      \cr
		         b  		      \cr
		         c  		      \cr} \right)
      = \left(\matrix{H_{11}&H_{12}&H_{13} \cr
                      H_{21}&H_{22}&H_{23} \cr
                      H_{31}&H_{32}&H_{33} \cr} \right)
           \left(\matrix{a  		      \cr
		         b  		      \cr
		         c  		      \cr} \right),
$$
where
$$\eqalign{
	H_{11}=&\left(m^2-\frac{N^2-1}{2N}\frac{g^2}{\pi}\right)A^{(2)}
	                 -\frac{N^2-1}{2N}\frac{g^2}{\pi}A^{(3)},
	\cr
	H_{22}=&2\left(m^2-\frac{N^2-1}{2N}\frac{g^2}{\pi}\right)
	        \left(B^{(2)}+B^{(3)}\right)
	                +\frac{N-1}{N}\frac{g^2}{\pi}B^{(4)}
	\cr
	       &-\frac{(N-1)(N+2)}{N}\frac{g^2}{\pi}B^{(5)}
	        +\frac{N-1}{2N}\frac{g^2}{\pi}\left(B^{(6)}+B^{(7)}\right),
	\cr
	H_{33}=&2\left(m^2-\frac{N^2-1}{2N}\frac{g^2}{\pi}\right)
	        \left(B^{(2)}+B^{(3)}\right)
	                +\frac{N+1}{N}\frac{g^2}{\pi}B^{(4)}
	\cr
	       &-\frac{(N+1)(N-2)}{N}\frac{g^2}{\pi}B^{(5)}
	        -\frac{N+1}{2N}\frac{g^2}{\pi}\left(B^{(6)}+B^{(7)}\right),
	\cr
	H_{12}=&H_{21}^{T}
	      =\frac{(N-1)\sqrt{2(N+1)}}{2N}\frac{g^2}{\pi}
	       \left(C^{(1)}-C^{(2)}\right),\hskip 6 em
	\cr
	H_{13}=&H_{31}^{T}
	      =\frac{(N+1)\sqrt{2(N-1)}}{2N}\frac{g^2}{\pi}
	       \left(C^{(1)}-C^{(2)}\right),\hskip 6 em
	\cr
	H_{23}=&H_{32}^{T}
	      =-\frac{\sqrt{(N+1)(N-1)}}{N}\frac{g^2}{\pi}
	       \left(B^{(4)}+NB^{(5)}\right),
	\cr                                  }\eqno{\rm (C.1)}
$$
\vfill\eject
$$\eqalign{
	{A^{(1)}}_{kl}&=\int_0^1 dx f_k(x)f_l(x),\cr
	{A^{(2)}}_{kl}&=\int_0^1 dx \frac{f_k(x)f_l(x)}{x(1-x)},\cr
	{A^{(3)}}_{kl}&=\int_0^1 dx dy \frac{f_k(x)f_l(y)}{(x-y)^2},\cr
        {B^{(1)}}_{kl}
	&=\int_{(4)} G_k(x_1,x_2,x_3,x_4)G_l(x_1,x_2,x_3,x_4),\cr
	{B^{(2)}}_{kl}
	&=\int_{(4)} G_k(x_1,x_2,x_3,x_4) \frac{1}{x_1}
	            G_l(x_1,x_2,x_3,x_4),\cr
	{B^{(3)}}_{kl}
	&=\int_{(4)} G_k(x_1,x_2,x_3,x_4) \frac{1}{x_2}
	            G_l(x_1,x_2,x_3,x_4),\cr
	{B^{(4)}}_{kl}
	&=\int_{(6)} G_k(x_1,x_2,x_3,x_4) \frac{1}{(x_1+x_2)^2}
	            G_l(y_1,y_2,x_3,x_4),\cr
	{B^{(5)}}_{kl}
	&=\int_{(6)} G_k(x_1,x_2,x_3,x_4) \frac{1}{(x_1-y_1)^2}
	            G_l(y_1,y_2,x_3,x_4),\cr
	{B^{(6)}}_{kl}
	&=\int_{(6)^{'}} G_k(x_1,x_2,x_3,x_4) \frac{1}{(x_1-y_1)^2}
	            G_l(y_1,x_2,y_2,x_4),\cr
	{B^{(7)}}_{kl}
	&=\int_{(6)^{''}} G_k(x_1,x_2,x_3,x_4) \frac{1}{(x_2-y_1)^2}
	            G_l(x_1,y_1,x_3,y_2),\cr
	{C^{(1)}}_{kl}
	&=\int_{(4)} f_k(x_1)\frac{1}{(x_2+x_3)^2}
	            G_l(x_1,x_2,x_3,x_4),\cr
	{C^{(2)}}_{kl}
	&=\int_{(4)} f_k(1-x_4)\frac{1}{(x_2+x_3)^2}
	            G_l(x_1,x_2,x_3,x_4),\cr  }    \eqno{\rm (C.2)}
$$
$$\eqalign{
	\int_{(4)} &\equiv \int_0^1 \prod_{i=1}^4 dx_i
	                  \delta (\sum_{i=1}^4 x_i -1),\cr
	\int_{(6)} &\equiv \int_0^1 \prod_{i=1}^4 dx_i dy_1 dy_2
	                  \delta (\sum_{i=1}^4 x_i -1)
	                  \delta (x_1+x_2-y_1-y_2),\cr
	\int_{(6)^{'}} &\equiv \int_0^1 \prod_{i=1}^4 dx_i dy_1 dy_2
	                  \delta (\sum_{i=1}^4 x_i -1)
	                  \delta (x_1+x_3-y_1-y_2),\cr
	\int_{(6)^{''}} &\equiv \int_0^1 \prod_{i=1}^4 dx_i dy_1 dy_2
	                  \delta (\sum_{i=1}^4 x_i -1)
	                  \delta (x_2+x_4-y_1-y_2).\cr  }\eqno{\rm (C.3)}
$$
These integrals with no $N$-dependence can be calculated analytically
with the formulae collected in Ref. [14].

%\vfill\eject

\centerline{\bf References}

\vskip 0.5 true cm
\item{[1]}  G. 't Hooft, Nucl. Phys. {\bf B75}(1974)461.
\item{[2]}  P. Steinhardt, Nucl. Phys. {\bf B176}(1980)100.
\item{[3]}  C.J. Hamer, Nucl. Phys. {\bf B121}(1977)159;
         ibid. {\bf B132}(1978)542.
\item{[4]}  T. Eller, H.C. Pauli and S. Brodsky, Phys. Rev. {\bf D35}(1987)
         1493.
\item{[5]}  T. Eller and H.C. Pauli, Z. Phys. {\bf C42}(1989) 59.
\item{[6]}  K. Hornbostel,S. Brodsky and H.Pauli,
         Phys. Rev. {\bf D41}(1990)3814.
\item{[7]}  F.Lentz,M.Thies,S.Levit and K.Yazaki,
         Ann. Phys. {\bf 208}(1991)1.
\item{[8]}  C.M. Yung and C.J. Hamer, Phys. Rev. {\bf D44}(1991) 2598.
\item{[9]}  R.J. Perry, A. Harindranath,and K.G. Wilson,
         Phys. Rev. Lett.{\bf65}(1990)2959.
\item{[10]}  I. Tamm, J. Phys. (USSR) {\bf 9}(1945)449;
         S.M. Dancoff, Phys. Rev. {\bf 78}(1950) 382;
         H.A. Bethe and F.D. Hoffman, {\it Mesons and Fields}
         (Row, Peterson,\break\hfill Evanston, 1955) Vol.{\Roman 2};
         E.M. Henley and W. Thirring,
         {\it Elementary Quantum Field Theory}
         (McGraw-Hill, New York,1962).
\item{[11]}  An extensive list of references on light-front physics
         by A.Harindranath (light. tex) is available via anonymous ftp
         from public. mps. ohio-state.edu under the subdirectory
         tmp/infolight.
\item{[12]}  Y. Mo and R.J. Perry, J. Comp. Phys. {\bf 108}(1993)159.
\item{[13]}  S.Coleman, Commun. Math. Phys. {\bf 31}(1973)259.
\item{[14]}  K. Harada, T. Sugihara, M. Taniguchi and M. Yahiro,
         Kyushu Univ. preprint KYUSHU-HET-9,
         Phys. Rev. {\bf D}, to be published.
\item{[15]}  G. 't Hooft, Nucl.Phys. {\bf B72}(1974)461.
\item{[16]}  E. Witten, Nucl. Phys. {\bf B160}(1979)57.
\item{[17]}  S.Coleman, {\it Aspects of Symmetry}
         (Cambridge, London, 1985).
\item{[18]}  R.J. Perry and A. Harindranath, Phys. Rev.
         {\bf D43}(1991) 4051.
\item{[19]}  W.M. Zhang and A. Harindranath, Phys. Rev.
         {\bf D48}(1993) 4868.
\item{[20]}  H. Bergknoff, Nucl. Phys. {\bf B122}(1977)215.
\vfill\eject
%
%   TABLE 1
%
\noindent
TABLE I. Calculated masses of the SU(2) and SU(3) hadronic bound states
$M_1$, $M_2$ and $M_{\rm b}$ are tabulated for various values of the quark
mass $m$.
Here all the masses are given in units of $\sqrt{g^2N/2\pi}$.

$$
\vbox{
\offinterlineskip
\halign{\strut
\vrule# & \quad\hfil#\hfil\quad & \vrule# &
          \quad\hfil#\hfil\quad & \vrule# &
          \quad\hfil#\hfil\quad & \vrule# \cr
\noalign{\hrule}
&       &&          $N=2$               &&          $N=3$              & \cr
\noalign{\hrule}
& $m$   && $M_1$ \hskip 1.0 true cm $M_2$ \hskip 1.0 true cm $M_b$
        && $M_1$ \hskip 1.0 true cm $M_2$ \hskip 1.0 true cm $M_b$ & \cr
\noalign{\hrule}
&0.0001 &&  0.01626 \ 0.02880 \ 0.01773 && 0.01849 \ 0.03666 \ 0.03203 & \cr
&0.0005 &&  0.03635 \ 0.06414 \ 0.03964 && 0.04134 \ 0.08142 \ 0.07162 & \cr
&0.0010 &&  0.05141 \ 0.09067 \ 0.05608 && 0.05846 \ 0.11495 \ 0.10130 & \cr
&0.0050 &&  0.11512 \ 0.20325 \ 0.12569 && 0.13053 \ 0.25674 \ 0.22660 & \cr
&0.0100 &&  0.16317 \ 0.28858 \ 0.17825 && 0.18443 \ 0.36336 \ 0.32062 & \cr
&0.0500 &&  0.37414 \ 0.66580 \ 0.40748 && 0.41486 \ 0.81963 \ 0.72106 & \cr
&0.1000 &&  0.54915 \ 0.97461 \ 0.59164 && 0.59895 \ 1.17476 \ 1.03014 & \cr
\noalign{\hrule}
 }
 }
$$
\vfill\eject
\centerline{\bf Figure Captions}
\vskip 0.5 true cm
\noindent
FIG.1
Masses of the lightest ($M_1$) and the second lightest ($M_2$) SU(2) mesons
are shown as a function of one of the parameters which characterize the size of
the space spanned by the basis functions (see eqs. (3.1) - (3.3));
(a) $N_2$ is varied while $N_4$ is fixed, (b) $N_4$ is varied while $N_2$ is
fixed.
Here the masses are presented in units of $\sqrt{g^2N/2\pi}$.
\vskip 0.5 true cm

\noindent
FIG.2
Masses of the lowest two mesonic ($M_1$ and $M_2$) and one baryonic
 ($M_{\rm b}$) bound
states, obtained with the full-fledged calculations, are shown in units of
$\sqrt{g^2N/2\pi}$ as a function of the quark mass $m$;
(a) for SU(2) and (b) for SU(3).
They are graphed with solid lines.
The 2-body decay threshold, 2$M_1$, is also shown by the dashed line.
\vskip 0.5 true cm

\noindent
FIG.3
Numerical and approximate masses of the lowest mesonic ($M_1$) and baryonic
($M_{\rm b}$) states are shown in units of $\sqrt{g^2N/2\pi}$ as a function
of the quark mass $m$.
(a) For the SU(2) case, the numerical solutions to $M_1$ and $M_{\rm b}$ are
graphed with the solid and dot-dashed lines, respectively, while the
approximate ones to both of them are degenerate and therefore graphed with
a dashed line.
(b) For the SU(3) case, the numerical and the approximate solutions are
graphed with the solid and dashed lines, respectively, for both $M_1$ and
$M_{\rm b}$.
\vskip 0.5 true cm

\noindent
FIG.4
Numerical solutions to (a) the second lightest mesonic and (b) the lightest
baryonic masses, calculated for SU($N$) by adopting a quark mass $m=10^{-4}$,
are shown by the solid lines as a function of $N$ in comparison to that of
the lightest meson.
The approximate solutions, which are available only for the lightest meson
and baryon, are shown by the dashed lines in (a) and (b), respectively.
The 2-body decay threshold, 2$M_1$, is also shown by the dot-dashed line in
 (a).
Here all the masses are given in units of $\sqrt{g^2N/2\pi}$.
\vskip 0.5 true cm

\noindent
FIG.5
Masses of the lowest four 2-body and a 4-body SU($N$) mesonic states in
the large $N$ limit are shown as a function of the quark mass $m$ by the
solid lines and the dot-dashed line, respectively.
The 2-body decay threshold, 2$M_1$, is also shown by the dashed line.
Note that all the masses are given in units of $\sqrt{g^2N/2\pi}$.

\bye